\documentclass[accepted=2019-02-13,a4paper,twocolumn,superscriptaddress,11pt]{quantumarticle}
\usepackage[numbers,sort&compress]{natbib}
\usepackage[T1]{fontenc}
\usepackage[latin9]{inputenc}
\usepackage{color}
\usepackage{array}
\usepackage{rotating}
\usepackage{float}
\usepackage{multirow}
\usepackage{graphicx}
\usepackage[unicode=true]{hyperref}

\makeatletter

\providecommand{\tabularnewline}{\\}

\usepackage{hyperref}
\usepackage[table]{xcolor}
\usepackage{ltablex}

\makeatother

\begin{document}

\title{Optimized Entanglement Purification}

\author{Stefan Krastanov}

\affiliation{Departments of Applied Physics and Physics, Yale University, New
Haven, CT 06511, USA}

\affiliation{Yale Quantum Institute, Yale University, New Haven, CT 06520, USA}

\author{Victor V. Albert}

\affiliation{Departments of Applied Physics and Physics, Yale University, New
Haven, CT 06511, USA}

\affiliation{Yale Quantum Institute, Yale University, New Haven, CT 06520, USA}

\affiliation{Walter Burke Institute for Theoretical Physics and Institute for
Quantum Information and Matter, California Institute of Technology,
Pasadena, California 91125, USA}

\author{Liang Jiang}

\affiliation{Departments of Applied Physics and Physics, Yale University, New
Haven, CT 06511, USA}

\affiliation{Yale Quantum Institute, Yale University, New Haven, CT 06520, USA}
\begin{abstract}
We investigate novel protocols for entanglement purification of qubit
Bell pairs. Employing genetic algorithms for the design
of the purification circuit, we obtain shorter circuits achieving
higher success rates and better final fidelities than what is currently
available in the literature. We provide a software tool for analytical
and numerical study of the generated purification circuits, under
customizable error models. These new purification protocols pave the
way to practical implementations of modular quantum computers and
quantum repeaters. Our approach is particularly attentive to the effects
of finite resources and imperfect local operations - phenomena neglected
in the usual asymptotic approach to the problem. The choice of the building
blocks permitted in the construction of the circuits is based on a thorough
enumeration of the local Clifford operations that act as permutations on the
basis of Bell states.
\end{abstract}

\maketitle

The eventual construction of a scalable quantum computer is bound
to revolutionize both how we solve practical problems like quantum
simulation, and how we approach foundational questions ranging from
topics in computational complexity to quantum gravity. However, numerous
engineering hurdles have to be surmounted along the way, as exemplified
by today's race to implement practical quantum error-correcting codes.
While great many high performing error-correcting codes have been
constructed by theorists, only recently did experiments start approaching
hardware-level error rates that are sufficiently close to the threshold
at which codes actually start to help~\citep{cramer2016repeated,ofek2016extending}.
A promising approach is the modular architecture~\citep{gottesman1999demonstrating,monroe2014large}
for quantum computers with implementations based on, among others, superconducting circuits~\citep{narla2016robust},
trapped ions~\citep{hucul2015modular,nigmatullin2016minimally}, or
NV centers~\citep{reiserer2016robust}.
The central theme is the creation of a network of small independent
quantum registers of few qubits, with connections capable of distributing
entangled pairs between nodes~\citep{nickerson2013topological,monroe2014large}.
Such an architecture avoids the difficulty of creating a single complex
structure as described in more monolithic approaches and offers a
systematic way to minimize undesired crosstalk and residual interactions
while scaling the system. Moreover, the same modules might also be
used for the design of quantum repeaters for use in quantum communication~\citep{dur1999quantum,pan2001entanglement,childress2005fault,jiang2007fast,nickerson2014freely}.

Experimentally, there have been significant advances in creating entanglement
between modules, with demonstrations in trapped ions~\citep{moehring2007entanglement,hucul2015modular},
NV centers~\citep{pfaff2013demonstration,hensen2015loophole}, neutral
atoms~\citep{ritter2012elementary}, and superconducting circuits~\citep{narla2016robust}.
However, the infidelity of created Bell pairs is on the order of 10\%,
while noise due to local gates and measurements can be much lower
than $1\%$. Purification of the entanglement resource will be necessary
before successfully employing it for fault-tolerant computation or
communication. Although various purification protocols have been proposed~\citep{deutsch1996quantum,bennett1996purification,dur1999quantum,dur2007entanglement,fujii2009entanglement,nickerson2013topological,nickerson2014freely,nigmatullin2016minimally},
there is still a lack of systematic comparison and optimization of
purification circuits, as the number of possible designs increases
exponentially with the size of the circuits. In this paper we develop
tools to generate and compare purification circuits and we present
multiple purification protocols outperforming the
contenders we test against~\citep{nickerson2013topological,nickerson2014freely,nigmatullin2016minimally}
over a wide range of realistic hardware parameters. We review the
notion of an entanglement purification circuit and present our approach
to generating and evaluating such circuits. We compare our results
to recent proposals for practical high-performance purification circuits,
and finally discuss the design principles and key ingredients for
efficient purification circuits.

Importantly, we pay particular attention to the limitations imposed
by working with finite hardware resources. One can find many highly efficient
purification schemes in the literature, which reach perfect fidelities
at high yield in the asymptotic regime (e.g. \citep{bennett1996purification,dur1999quantum}),
however such asymptotic resource theories neglect the imperfections and size limitations
of the purification hardware. Moreover, a large family of such circuits can be
constructed from error correcting
codes\citep{bennett1996purification,aschauer2005thesis}, however they can often be
impractically wide as they do not exploit the possibility of renewed generation of
entanglement in already measured qubit registers.
Our work optimizes entanglement purification in today's NISQ~\citep{preskill2018nisq} devices, complementing the protocols that are optimal only in the asymptotic regime of arbitrarily many available qubits and perfect gates and measurements.
The imperfections in the local gates
and measurements are the limiting factor in real-world hardware.
We compare our results to other known finite
protocols (including Oxfords's and IBM's~\citep{deutsch1996quantum,bennett1996mixed}, STRINGENT~\citep{nickerson2013topological,fujii2009entanglement},
and some recursive
or iterative~\citep{dur1999quantum} versions of the same).

\begin{figure}
\begin{center}
\includegraphics[width=3cm]{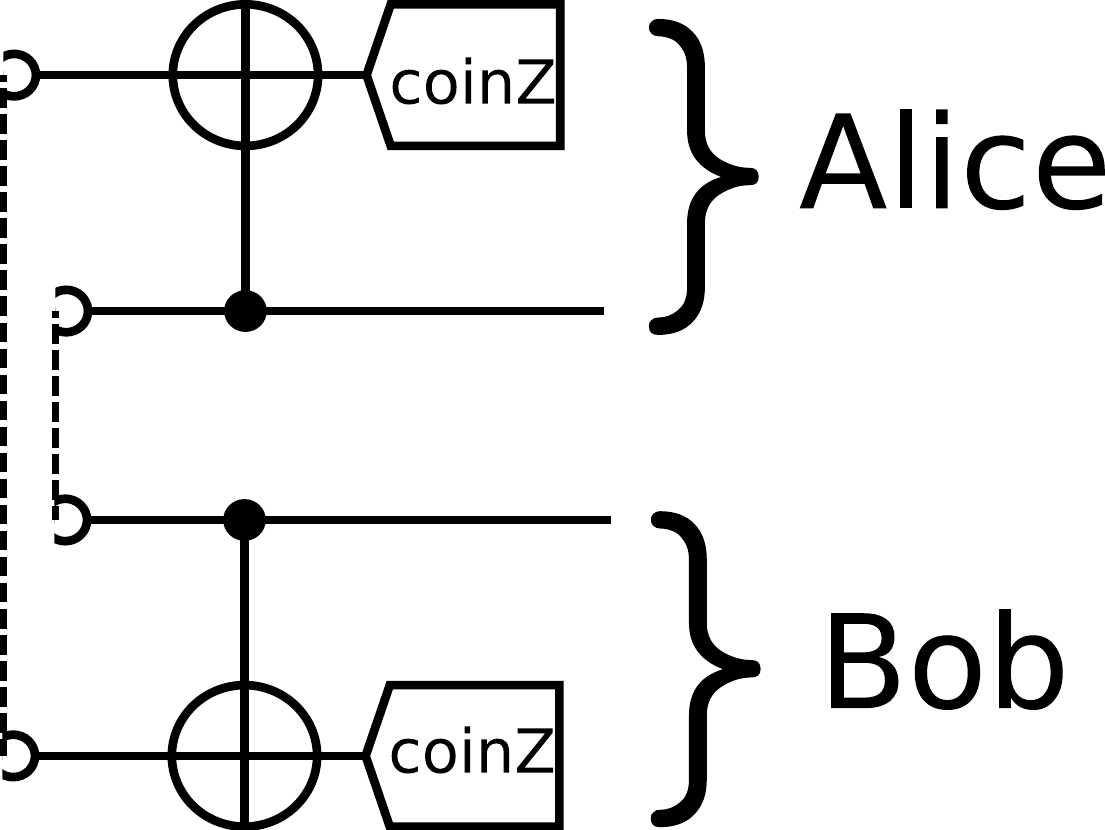}
\end{center}
\caption{\label{fig:simple-purification}A simple purification circuit of width
2 (i.e. 2 local qubits for Alice or Bob.). The upper half is ran by
Alice, while the bottom half is ran by Bob. The dashed lines correspond
to the initialization of registers with low-quality ``raw'' Bell
pairs. The top and bottom register correspond to the two qubits of
the sacrificial Bell pair. A coincidence measurement in the Z basis
marks a successful purification procedure.}
\end{figure}

\paragraph{\label{sec:Bell-Pairs-Purification}Purification of Bell Pairs}

In an entanglement purification protocol, two parties, Alice and Bob,
start by sharing a number of imperfect Bell pairs and by performing
local gates and measurements and communicating classically, they obtain
a single pair of higher fidelity. For conciseness we use A, B, C,
and D to denote the Bell basis states.
\begin{equation}
\begin{array}{c}
A=|\phi_{+}\rangle=\frac{|00\rangle+|11\rangle}{\sqrt{2}}\\
B=|\psi_{-}\rangle=\frac{|01\rangle-|10\rangle}{\sqrt{2}}\\
C=|\psi_{+}\rangle=\frac{|01\rangle+|10\rangle}{\sqrt{2}}\\
D=|\phi_{-}\rangle=\frac{|00\rangle-|11\rangle}{\sqrt{2}}
\end{array}\label{eq:Bell_basis}
\end{equation}
 The imperfect pairs are described in the Bell basis (eq.~\ref{eq:Bell_basis})
by $\rho_{0}=F_{0}|A\rangle\langle A|+\frac{1-F_{0}}{3}(|B\rangle\langle B|+|C\rangle\langle C|+|D\rangle\langle D|)$.
If we have a state of multiple pairs (like AA), the first letter will
denote the pair to be purified.

To explain the roles of the local gates and coincidence measurements
let us consider, in Fig.~\ref{fig:simple-purification}, the simplest
purification circuit, in which Alice and Bob share two Bell pairs
and sacrifice one of them. One way to explain
the circuit is to describe it as an error-detecting circuit: If we
start with two perfectly initialized Bell pairs in the state A, then
the coincidence measurement will always succeed; however, if an X
error (a bit flip) happens on one of the qubits, that error will be
propagated by the CNOT operations and it will cause the coincidence
measurement to fail (the two qubits will point in opposite directions
on the Z axis). It is important to note that only X and Y errors can
be detected by this circuit, but not Z errors (phase flips) as the
coincidence measurement can not distinguish A from D states. One needs
a circuit running on more than two Bell pairs to address X, Y, and
Z errors. 

For the purpose of designing an optimal purification circuit, it is
enlightening to also interpret the local operations in terms of permutations
of the basis vectors\citep{dehaene2003local,bombin2005entanglement}.
The initial state of the 4 qubit system is described by the density
matrix $\rho_{0}\otimes\rho_{0}$, or equivalently by the 16 scalars
in its diagonal in the Bell basis $\{AA,AB,\dots,DD\}$. The ``mirrored''
CNOT operations that both Alice and Bob perform result in a new diagonal
density matrix with diagonal entries being a permutation of those
of the original density matrix. A coincidence measurement on the Z
axis follows, which results in projecting out half of the possible
states, i.e. deleting 8 of the scalars and renormalizing and adding
by pairs the other 8. The permutation operation and coincidence measurement
have to be chosen together such that this projection (when the coincidence
measurement is successful) results in filtering out many of the lower-probability
B, C, and D states. A detailed run through this example is given in
the supplementary materials.

If we restrict ourselves to finding the best ``single sacrifice''
circuits, i.e. circuits that sacrifice one Bell pair in an attempt
to purify another one, we need to find the best set of permutations
and measurements. There are 3 coincidence measurements of interest
- coincidence in the Z basis which selects for A and D; coincidence
in X basis which selects for A and C; and anti-coincidence in the
Y basis which selects for A and B. All of those measurements can be
implemented as a Z measurement preceded by a local Clifford operation.

The group of possible permutations is rather complicated. Firstly,
all permutations of the Bell basis are Clifford operations because
the permutation operation can be written as a permutation on the stabilizers
of each state (moreover, we do not have access to all 16! permutations,
as only operations local for Alice and Bob are permitted). This restriction
permits us to efficiently enumerate all possible permutations and
study their performance. The software for performing this enumeration
is provided with this manuscript. The enumeration goes as follows~\citep{ozols2008clifford,calderbank1998quantum}.
There are 11520 operations in the Clifford group of two qubits $C_{2}$.
After exhaustively listing all operations in $C_{2}^{\otimes2}$ we
are left with 184320 unique Clifford operations that act as permutations
of the Bell basis of two Bell pairs. Accounting for 16 different operations
that change only the global phase of the state (e.g. XX which maps
B to -B) we are left with 11520 unique permutations. Restricting ourselves
to permutations that map A to A cuts that number by a factor of 4
for each pair, which leaves us with 720 unique restricted permutations.
Out of those, 72 operations do not change the fidelity ($72=2\times6\times6$
correspond to two operations (the identity and SWAP) under the six
possible BCD permutations for each pair). The remaining 648 permitted
operations perform equally well when purifying against depolarization
noise, if they are used with the appropriate coincidence measurement.
Half of them can be generated from the mirrored CNOT operation from
Fig.~\ref{fig:simple-purification} together with BCD permutations
performed before or after on each of the two pairs. The other half
can be generated if we also employ a SWAP gate (such gate can be of
importance for hardware implementations that have ``hot'' communication
qubits and ``cold'' storage qubits\citep{nigmatullin2016minimally}).
When we break the symmetry of the depolarization noise and use biased
noise instead, all of these operations still permit purification,
but a small fraction of them significantly outperform the rest. So
far, we have only counted purification circuits with 2 local qubits.
We may increase the width (i.e., number of local qubits) to boost
the performance of the entanglement purification. However, the number
of possible purification circuits grows exponentially with not only
the length, but also the width of the circuit. Even for relatively
small circuits (e.g., length 10 and width 3), there will be $>10^{40}$
different configurations, if we use the operations discussed above,
which are impossible to exhaustively compare. Therefore, we need an
efficient procedure to choose the appropriate permutation operation
at each round of our purification protocol.

\begin{figure}[H]
\includegraphics[width=7.5cm]{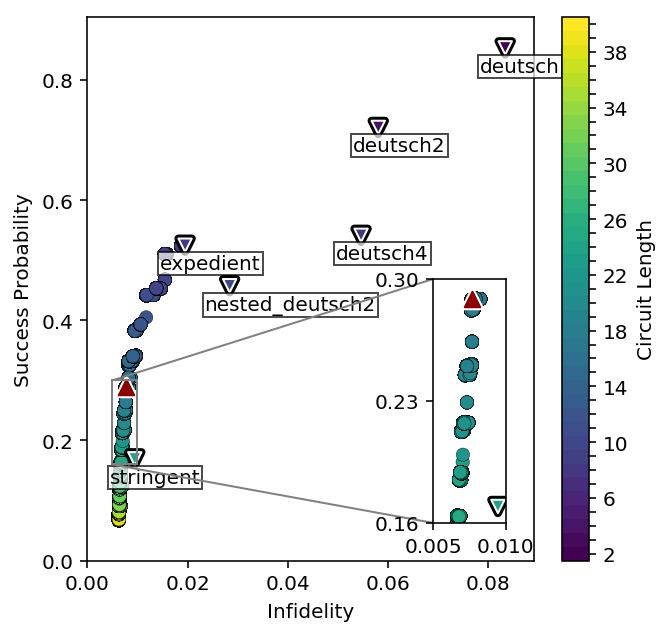}\caption{\label{fig:comparing-circuits}Comparing circuits designed by
our genetic algorithm (for three-qubit registers) to prior art. Each circle marks a unique circuit. The
horizontal axis is the infidelity of the final pair.
The vertical axis is the probability of success. Also shown are
the ``Oxford'' scheme on two Bell pairs~\citep{deutsch1996quantum},
which outperforms the IBM scheme~\citep{bennett1996purification}.
The Innsbruck's ``Pumping(2)'' and ``Pumping(4)'' are that same scheme
applied consecutively two or four times from~\citep{dur1999quantum}(Sec~3.a).
The aforementioned schemes require two-qubit registers while the rest are for three-qubit registers.
 ``Recurrent(2)'' is the recursive version (at depth 2) of ``Oxford'' from\citep{bennett1996purification,dur1999quantum}.
In ``Rec.\&Pump.'' one recursively repeats the pumping protocol instead of the Oxford protocol. To our knowledge
``EXPEDIENT'' and ``STRINGENT'' are some of the best circuits \citep{nickerson2013topological,fujii2009entanglement}).
Evaluations done at $p_{2}=\eta=0.99$ and $F_{0}=0.9$. The red triangle marks a
circuit of ours we discuss more in Fig.~\ref{fig:L17-STRINGENT}.}
\end{figure}

\paragraph{Discrete optimization algorithm}

The design of circuits, whether quantum or classical, lends itself
naturally to the use of evolutionary (or genetic) algorithms with numerous interesting
examples in, for instance, electronics and robotics~\citep{geijtenbeek2013flexible}.
In particular, in the field of quantum information, such optimization techniques have been used in widely different settings, including
control~\citep{weidner2017atom}, state preparation
and metrology~\citep{knott2016metrology}, and studies of locality as related to Bell's inequality~\citep{harper2017correlations}.
An evolutionary algorithm is an optimization algorithm particularly
useful for cost functions over discrete parameter spaces. A candidate
solution (a point in parameter space) plays the role of an individual
in a population subjected to simulated evolution. Depending on the
particular implementation, each individual generates a number of children,
whether through mutations or through sexual modes of reproduction
with other individuals in the population. The population is then culled
so only the fittest candidate-solutions remain and the procedure is
repeated for multiple generations until the convergence criteria are
fulfilled.

In our particular implementation\footnote{https://qevo.krastanov.org/\ (online repository)\ The provided
software tools and examples in the supplementary materials are readily
usable in pedagogical settings as well. Moreover, our software provides
analytical expressions for the final fidelities and numerical estimates
for the expected resource overhead. The circuits can be fine-tuned
during the optimization run for the error model of the particular
hardware. This online resource is cloned at krastanov.github.io/qevo/index.html.} the individuals are quantum entanglement purification circuits. We
restrict ourselves to circuits that purify the entanglement between
two parties, Alice and Bob, without the involvement of a third party.
The circuit can contain any of the previously discussed coincidence
measurements (coincidence in Z, coincidence in X, anti-coincidence
in Y). The circuit is also permitted to contain the ``mirrored''
CNOT operation from Fig.~\ref{fig:simple-purification} together
with any permutation of the \{B,C,D\} states applied before the CNOT
operation. Applying the \{B,C,D\} permutation after the CNOT is unnecessary
as the next operation would already have that degree of freedom. However,
the final result will have a biased error, so a single \{B,C,D\} permutation
at the very end might be required.

\paragraph{Operation and measurement errors}

The design of the purification protocol is sensitive to imperfections
in the local operations as well. We parameterize the operational infidelities
with the parameters $p_{2}$, where $1-p_{2}$ is the chance for a
two-qubit gate to cause a depolarization, and $\eta$, where $1-\eta$
is the chance for a measurement to report the incorrect result. 

No memory errors or single-qubit gate errors are considered in our
treatment as they are generally much smaller~\citep{barends2014superconducting,ballance2016high},
but they can be accounted for in the same manner.

After a measurement the measured qubit pair is reset to a new Bell
pair and Alice and Bob can again use it as a purification resource.
The initial fidelity of each Bell pair is a parameter $F_{0}$ set
at the beginning of the algorithm. Similarly, we set the measurement
fidelity $\eta$ and the two-qubit gate fidelity $p_{2}$ at the beginning.
More complicated settings with different error models are possible
as well \textendash{} of special interest would be circuits adapted
for registers containing a ``hot'' communication side (e.g. only one qubit in the register is able to establish initial remote entanglement) and a ``cold''
memory side (considered in Nigmatullin et al.~\citep{nigmatullin2016minimally}). For that type of registers one would also
add the SWAP gate to the permitted genome. However, we show such circuits only in the supplementary
materials as the majority of circuits in the literature
are designed for registers where all qubits have similar properties.

The ``fitness'' that we optimize is the fidelity of the final purified
Bell pair, however different weights can be placed on the ``infidelity
components'' along $|B\rangle$, $|C\rangle$, and $|D\rangle$ if
needed. In practice, the genetic algorithm is fairly robust to changing
parameters like the population size, mutation rates, or number of
children circuits. We provide both pregenerated circuits and ready-to-run
scripts to generate circuits from scratch.

Importantly, depending on the parameter regime and error model, different
circuits would be the top performers. This showcases the importance
of rerunning the optimization algorithm for the given hardware. An
example of such difference is provided in the supplementary materials.

\paragraph*{A word of caution: purification yield and finite imperfect circuits}

A common way to evaluate the performance of a purification circuit is to present its yield, defined as follows.
For a circuit that starts with $n$ imperfect Bell pairs and that produces $m$ Bell pairs with fidelity arbitrarily close to 1, the
yield is $\lim_{n \to \infty}\frac{m}{n}$. The limiting procedure is necessary due to the requirement that the final fidelity
needs to be arbitrarily close to 1, which is impossible for a finite circuit.
In some cases this can be achieved by the recursive (nested) application of a known finite circuit
or by the use of the more advanced hashing method~\citep{bennett1996mixed}, as long as the local operations and measurements
employed in the circuit are perfect.

However, our focus is on finite circuits of practical interest for near-term hardware. The yield is not a good measure of efficiency in
this case as it is defined in terms of a limiting procedure for asymptotic circuits. Moreover, we
specifically optimize our circuits to work in the presence of measurement and operational errors, which makes 
unit fidelities (required for the definition of yield) impossible for a finite circuit. Optimizing in the presence
of local errors also makes our circuits better than circuits that would have been optimized for a figure of merit that
neglects such errors (the supplementary materials illustrate this with examples of circuits designed for different levels of errors).

We use the final fidelity, the probability of success or the average amount of consumed raw Bell pairs as
measurements of efficiency, as these are the quantities of interest in the implementation of small error correcting codes on modular
architectures~\citep{nickerson2013topological} (these quantities decide the delay necessary for the performance of a stabilizer
measurement or gate teleportation). 

There is an interesting definition of yield that can be used for finite circuits like ours.
If one concatenates a finite circuit with the hashing method for purification one can use the yield of the
new asymptotic circuit as a figure of merit for the initial finite circuit. This "hashing yield" is still only
defined in the presence of perfect measurements and local operations, hence we do not employ it in the main text. However, we discuss it
in the supplementary materials.

\begin{figure*}
\includegraphics[width=16cm]{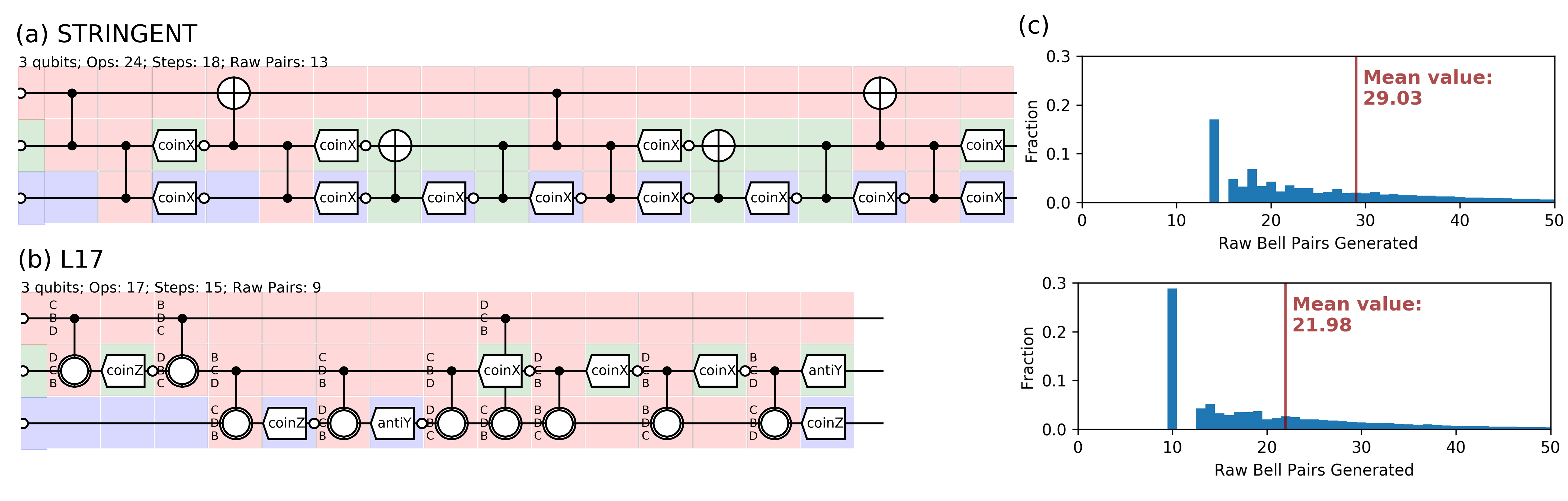}

\caption{\label{fig:L17-STRINGENT}Comparison of (a) the STRINGENT circuit~\citep{nickerson2013topological}
to (b) the L17 circuit obtained through optimization. The L17 circuit
outperforms STRINGENT in terms of both final obtained fidelity and
success probability over a wide range of error parameters. The color
coding shows independent sub-circuits in the STRINGENT circuit and
no such sub-circuits in our design. We show only Bob's side of the
circuit. The vertical set of letters before each gate marks how the
\{B,C,D\} states are permuted, which can be achieved with single qubit
gates folded in the CNOT gate as described in the supplementary materials.
While we use only CNOT two-qubit gates, we intentionally used a modified
symbol in order to bring attention to the presence of these permutation
operations. The small white circles after each measurement represent
generation of a new Bell pair resource with fidelity $F_{0}$. The
histograms (c) are of the required number of Bell pairs for a completion
of the protocol (as opposed to a single-shot run) for STRINGENT and the optimized
circuit. We also provide analytical evaluations of all our circuits at our online repository. In particular
even though STRINGENT is longer (24 operations single shot, 29 raw Bell pairs on average) than L17 (17 operations single shot, 22 raw Bell pairs on average), STRINGENT has higher final infidelity
by a factor of $\frac{16}{6}\approx 1.33$ if there are no measurement and operational errors. If we include them, they become
the dominant contribution to the final infidelity and in the case of STRINGENT vs L17 the infidelity is still higher by a factor of
1.26 (infidelities of 0.95\% and 0.77\%, success probabilities of 17\% and 29\%, respectively). The optimization and evaluation was done for $F_{0}=0.9$, $p_{2}=\eta=0.99$, without
memory errors or errors from single-qubit unitaries.}
\end{figure*}

Lastly, one can relax the requirement in the definition of yield that the final fidelity has to be 1. In that case
two new definitions of yield can be considered: $1/N$ where $N$ is the number of raw Bell pairs that the circuit
would require in a best case run ($N$ is represented as color in Fig.~\ref{fig:comparing-circuits}) or $1/N_{avg}$
where $N_{avg}$ is how many raw Bell pairs are expended on average until completion (taking into account that
the circuit might need to be restarted after a failed measurement). $N_{avg}$ is calculated through a Monte Carlo simulation
and also shown in Fig.~\ref{fig:L17-STRINGENT} for some of the circuits of interest. While $N$ depends only on the
particular circuit, $N_{avg}$ is a function of the error model (e.g. $F_{0}$, $p_{2}$, and $\eta$).

\paragraph{\label{sec:Comparing-to-Prior-Art}Comparing with Prior Art}

We generated a few thousand well performing circuits of lengths up
to 40 operations and acting on up to six pairs of qubits (and the
algorithm can easily generate bigger circuits, but soon one hits a
wall in performance due to the imperfections in the local operations
as discussed below). The zoo of circuits we have created can be explored
online (see note~1), but importantly, one can generate circuits specifically
for their hardware using our method. Fig.~\ref{fig:comparing-circuits}
shows how our circuits compare to a number of other circuits of width 3.
We outperform all circuits in the comparison in terms of final fidelity
of the distilled pair, while also having higher probabilities of success,
and employing fewer resources or shorter circuits. 

In Fig.~\ref{fig:L17-STRINGENT} we compare one of the best performing
circuits we know (STRINGENT from \citep{nickerson2013topological})
to one particular circuit we have designed. We show only Bob's side
of the circuit. Alice performs the same operations and the two parties
communicate to perform coincidence measurements. The shading permits
us to see which qubit pairs are engaged with other pairs: Each qubit
Bob possesses starts with a distinct color; The color is ``contagious'',
i.e. two-qubit gates will ``infect'' the second qubit in the gate
with the color of the first qubit; Measurements followed by regeneration
of the raw Bell pair resource reset the color of the measured qubit.
The shading clearly shows that the best protocols we find have all
the qubits engaged (entangled) together, a finding consistent with
the use of ``multiple selection'' purification protocols introduced
in \citep{fujii2009entanglement}. In contrast, conventional purification
protocols have sub-circuits where only a subset of the qubits are
engaged together.

A potential caveat of that ``completely engaged'' approach needs
to be addressed: in Fig.~\ref{fig:comparing-circuits} we report
the probability of all measurement in a given protocol succeeding
in a given run, but we do not report the following overhead. If a
single measurement fails, the protocol needs to be restarted; the
aforementioned conventional purification protocols, that posses sub-circuits,
can redo just the failed sub-circuit (for instance, either of the
two green blocks in STRINGENT from Fig.~\ref{fig:L17-STRINGENT}),
instead of restarting the entire circuit. A priori, this might lead
to lower resource overhead compared to our protocols (as they generally
completely entangle the qubits of each register), even if we still
win in terms of final fidelity and probability of success. However,
a detailed evaluation of this overhead shows that even when taken
into account, our protocols outperform the approaches we compare with as they require
lower number of gates, both in best case scenarios and on average
(see right panel of Fig.~\ref{fig:L17-STRINGENT}).

\begin{figure}
\includegraphics[width=8cm]{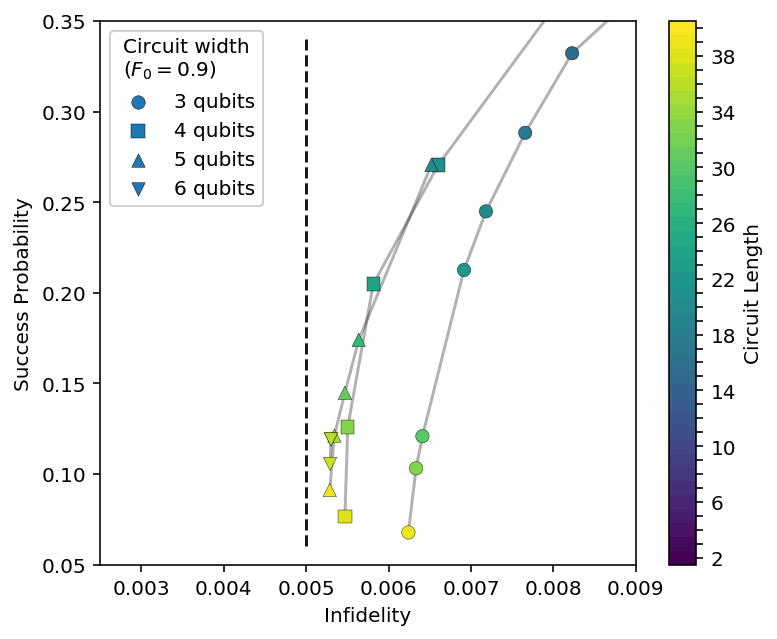}

\caption{\label{fig:N3vsN4}Similarly to Fig.~\ref{fig:comparing-circuits}
we compare the performance of circuits acting on 3 or more pairs.
For legibility, only some of the best generated circuits of each width
are shown (evaluated at $F_{0}=0.9$ and $p_{2}=\eta=0.99$). Our
circuits approach the limit of $\frac{\varepsilon}{2}=0.005$, derived
in the main text and supplementary materials. }
\end{figure}

Our approach can be employed for circuits with more than two sacrificial
pairs. Fig.~\ref{fig:N3vsN4} compares the performance of circuits
working on 3 pairs (as above) and circuits working on 4 or more pairs
of qubits. With still bigger circuits one quickly reaches a fidelity
limit imposed by the finite imperfections in the last operations performed
by the circuit (for hardware with perfect local operations that limitation
does not exist and one would rather use larger circuits that perform
many more operations per sacrifice~\citep{bennett1996mixed,dur2007entanglement,renes2015efficient}).
Example circuits are given online (see note~1). For much wider circuits
one can surpass this limitation and even use error correcting codes in order
to perform perfect (logical) local operations~\citep{zhou2016purification,zhou2017polarization},
but currently our software is not applicable to these cases.

\paragraph*{Operation versus initialization errors}

The design of efficient purification circuits needs to balance between
the initialization errors (imperfect raw Bell pairs) and operation
errors (imperfect local gates and measurements). As detailed in the
supplementary materials, for arbitrary long purification circuits,
the asymptotic infidelity is $\frac{\varepsilon}{2}+\mathcal{O}(\varepsilon^{2})$
where $\varepsilon=1-p_{2}$ (as indicated by the vertical dashed
line in Fig.~\ref{fig:N3vsN4}), which is only limited by the operation
errors. For finite-length purification circuits, however, the initialization
errors also play an important role, which determines how fast the
purification circuits approach the asymptotic limit with increasing
circuit length (Fig.~\ref{fig:N3vsN4}). By analyzing the circuits
given by the discrete optimization algorithm, we have observed that:
(1) For fixed length, depending on the parameter regime and error
model, different circuits would be the top performers. This showcases
the importance of rerunning the optimization algorithm for the given
hardware; (2) To boost the achievable fidelity, it is important to
use double-selection (where two Bell pairs are simultaneously sacrificed
to detect errors on a third surviving Bell pair)~\citep{fujii2009entanglement}
instead of repeated single-selection (where only one Bell pair is
sacrificed at each error detection step). This stems from the fact
that the asymptotic infidelity of single-selection is $\frac{7\varepsilon}{8}$,
i.e. nearly twice that of double-selection. Moreover, multiple selection
(where $n>2$ Bell pairs are simultaneously sacrificed) has the same
dominant asymptotic infidelity of $\frac{\varepsilon}{2}$ as double
selection. Therefore, wider circuits provide diminishing returns in the presence of operational errors. 

While the operational errors limit the utility of very wide circuits, in the case
of perfect local operations this phenomenon of "diminishing returns" does not exist.
In that case one can still use our optimization algorithm to design good small circuits,
but for large circuits the cost function as currently defined can not be computed efficiently.
Another possible direction of interest, which would require the simulation of large circuits, would be the
application of discrete optimization algorithms to the purification of multi-party entanglement~\citep{murao1998multiparticle,frowis2011stable,frowis2012stability,nickerson2013topological}, however this might be computationally prohibitive.
Of note is, however, that the computational complexity stems from the tracking of the purely classical
probabilities at each measurement branch (the circuits are composed of Clifford gates that can be
simulated efficiently). We are hopeful that stochastic evaluation of the cost function (through
a Monte Carlo method) will be sufficient to surmount any computational challenges in the application
of our protocol to larger circuits, however we do not pursue this in the current work.

In conclusion, we have optimized purification circuits of fixed width
using a discrete optimization approach using ``building-block''
subcircuits proven to be optimal. The optimized circuits outperform
many other general-purpose purification protocols in all three aspects
\textendash{} fidelity of purified Bell pair, success probability,
and circuit length (whether measured in terms of average number of
operations performed or average number of raw Bell pairs used). For purification circuits of width 2, we analyze the group
structure of the Clifford operations that fulfill the locality constraints
of purification. For purification circuits of width $\ge3$, we demonstrate
the importance of multiple selection (using at least two sacrificial
Bell pairs to simultaneously detect errors), and specify the diminishing
returns of using much wider circuits. We numerically obtain efficient
purification circuits that approach the asymptotic theoretical limits.
Our approach of using discrete optimization algorithms is applicable
to various errors models (e.g., dephasing dominated gate errors, imperfect
Bell state beyond the Werner form, etc). Moreover, it can be used
to optimize the purification circuits in the presence of memory errors,
including additional decoherence to all local qubits during the creation
of Bell pairs, and to investigate the entanglement purification of
encoded Bell pairs.
\begin{acknowledgments}
We are grateful for the helpful input from Holly Mandel and Kyungjoo Noh. This work would not have
been possible without the contributions of the Python, Jupyter, Matplotlib,
Numpy, Sympy, Qutip, and Julia open source projects and the Yale HPC
team. We acknowledge support from ARL-CDQI, ARO (W911NF-14-1-0011,
W911NF-14-1-0563), ARO MURI (W911NF-16-1-0349 ), AFOSR MURI (FA9550-14-1-0052,
FA9550-15-1-0015), the Alfred P. Sloan Foundation (BR2013-049), and
the Packard Foundation (2013-39273).
\end{acknowledgments}

\bibliographystyle{unsrtnat}
\bibliography{qevo}

\newpage{}

\appendix

\section*{Supplementary Materials}

The software and additional online materials are available at \href{http://qevo.krastanov.org}{qevo.krastanov.org} and krastanov.github.io/qevo/index.html. 

\section{Model for operational errors}

We consider each two-qubit gate $\hat{U}$ to be performed correctly
with a chance $p_{2}$ and to completely depolarize the two qubits
$i$ and $j$ it is acting upon with chance $1-p_{2}$. Written as
density matrices, when applied to input $\hat{\rho}_{in}$ it results
in 
\begin{equation}
\hat{\rho}_{out}=p_{2}\hat{U}\rho_{in}\hat{U}^{\dagger}+(1-p_{2})Tr_{i,j}(\hat{\rho}_{in})\otimes\frac{\hat{I}_{i,j}}{4},\label{eq:op_depolarize}
\end{equation}
where $Tr_{i,j}$ is a partial trace over the affected qubits and
$I_{i,j}$ is the identity operator associated with qubits $i$ and
$j$. Similarly, measurement on qubit $i$ has a probability $\eta$
to properly project and measure and a probability $1-\eta$ to erroneously
report the opposite result (flipping the qubit in the measurement
basis). For instance, an imperfect projection on $|1\rangle$ reads
as 
\begin{equation}
\hat{\rho}_{out}=\eta|1\rangle\langle1|\rho_{in}|1\rangle\langle1|+(1-\eta)|0\rangle\langle0|\hat{\rho}_{in}|0\rangle\langle0|.
\end{equation}

Memory errors are not considered, but can be easily added to the optimization
if required.

\section{Purification example when operations are interpreted as permutations
of the Bell basis}

Consider the simple circuit from Fig.~\ref{fig:simple-purification}.
As discussed throughout the main text, a useful way to represent the
operations performed in the circuit is as permutations of the Bell
basis. For this example we will use perfect operations (i.e. only
initialization errors). The density matrix describing the system will
be diagonal throughout the execution of the entire protocol as only
permutation operations are performed. The following table describes
how ``mirrored'' CNOT operation acts on the basis states (``AD''
stands for ``the sacrificial pair is in state D and the pair to be
purified is in state A''):
\begin{center}
\begin{tabular}{|c|c|}
\hline 
initial state & mapped to\tabularnewline
\hline 
\hline 
AA & AA\tabularnewline
\hline 
AB & DB\tabularnewline
\hline 
AC & AC\tabularnewline
\hline 
AD & DD\tabularnewline
\hline 
BA & BC\tabularnewline
\hline 
BB & CD\tabularnewline
\hline 
BC & BA\tabularnewline
\hline 
BD & CB\tabularnewline
\hline 
CA & CC\tabularnewline
\hline 
CB & BD\tabularnewline
\hline 
CC & CA\tabularnewline
\hline 
CD & BB\tabularnewline
\hline 
DA & DA\tabularnewline
\hline 
DB & AB\tabularnewline
\hline 
DC & DC\tabularnewline
\hline 
DD & AD\tabularnewline
\hline 
\end{tabular}
\par\end{center}

With this mapping we can trace how the state of the system evolves.
The following table gives the diagonal of the density matrix describing
the system at each step. In the table $q=\frac{1-F}{3}$. The measurement
column assumes a successful coincidence measurement has been performed.
By normalizing and tallying the states that remain in A (for the purified
pair) we are left with fidelity after purification $F_{final}=\frac{F^{2}+q^{2}}{F^{2}+5q^{2}+2Fq}>F$.

\begin{center}
\begin{tabular}{|c|c|c|c|}
\hline 
state & initial & after CNOT & final \tabularnewline
\hline 
\hline 
AA & $F^{2}$ & $F^{2}$ & $F^{2}$\tabularnewline
\hline 
AB & $Fq$ & $q^{2}$ & \tabularnewline
\hline 
AC & $Fq$ & $Fq$ & \tabularnewline
\hline 
AD & $Fq$ & $q^{2}$ & $q^{2}$\tabularnewline
\hline 
BA & $Fq$ & $q^{2}$ & $q^{2}$\tabularnewline
\hline 
BB & $q^{2}$ & $q^{2}$ & \tabularnewline
\hline 
BC & $q^{2}$ & $Fq$ & \tabularnewline
\hline 
BD & $q^{2}$ & $q^{2}$ & $q^{2}$\tabularnewline
\hline 
CA & $Fq$ & $q^{2}$ & $q^{2}$\tabularnewline
\hline 
CB & $q^{2}$ & $q^{2}$ & \tabularnewline
\hline 
CC & $q^{2}$ & $Fq$ & \tabularnewline
\hline 
CD & $q^{2}$ & $q^{2}$ & $q^{2}$\tabularnewline
\hline 
DA & $Fq$ & $Fq$ & $Fq$\tabularnewline
\hline 
DB & $q^{2}$ & $Fq$ & \tabularnewline
\hline 
DC & $q^{2}$ & $q^{2}$ & \tabularnewline
\hline 
DD & $q^{2}$ & $Fq$ & $Fq$\tabularnewline
\hline 
\end{tabular}
\end{center}

Table~\ref{tab:measurements} gives more details on how different
coincidence measurements act on the Bell pair.

\section{Purification example when operations are interpreted as error detections}

\begin{table}[H]
\begin{centering}
\begin{tabular}{|c|c|}
\hline 
\cellcolor{gray}\textcolor{white}{coinX} & A or C\tabularnewline
\hline 
antiX & B or D\tabularnewline
\hline 
\end{tabular} %
\begin{tabular}{|c|c|}
\hline 
coinY & C or D\tabularnewline
\hline 
\cellcolor{gray}\textcolor{white}{antiY} & A or B\tabularnewline
\hline 
\end{tabular} %
\begin{tabular}{|c|c|}
\hline 
\cellcolor{gray}\textcolor{white}{coinZ} & A or D\tabularnewline
\hline 
antiZ & B or C\tabularnewline
\hline 
\end{tabular}
\par\end{centering}
\begin{centering}
\begin{tabular}{|c|c|}
\hline 
coinX & detects Y and Z\tabularnewline
\hline 
antiY & detects X and Z\tabularnewline
\hline 
coinZ & detects X and Y\tabularnewline
\hline 
\end{tabular}
\par\end{centering}
\centering{}\caption{\label{tab:measurements}Coincidence and Anticoincidence Measurements.
The three tables at the top show which two Bell states are selected
by different Bell measurements. The measurements that select A, the
state we are distilling, are highlighted. The other 3 possible coincidence
measurements do not select for A so they are not highlighted, nor
used as building blocks for our circuits. For the 3 measurements preserving
A, the bottom table restates the same information in terms of what
single qubit errors (I, X, Y, or Z Pauli operations) are detectable
by each (if we have started in state A).}
\end{table}

Another way to interpret the purification protocols is to look at
them as error detection protocols. This way of thinking was used in
the main text in the discussion of the limits imposed by the operational
errors. Here we will repeat this discussion with more pedagogical
visual aids for a particular choice of two-qubit operation and measurement.
As in the main text, we will first consider a single selection circuit
(where Alice and Bob share two Bell pairs and sacrifice one of them
to detect errors on the other one). We are showing only Bob's side
of the circuit.
\begin{center}
\includegraphics[width=2cm]{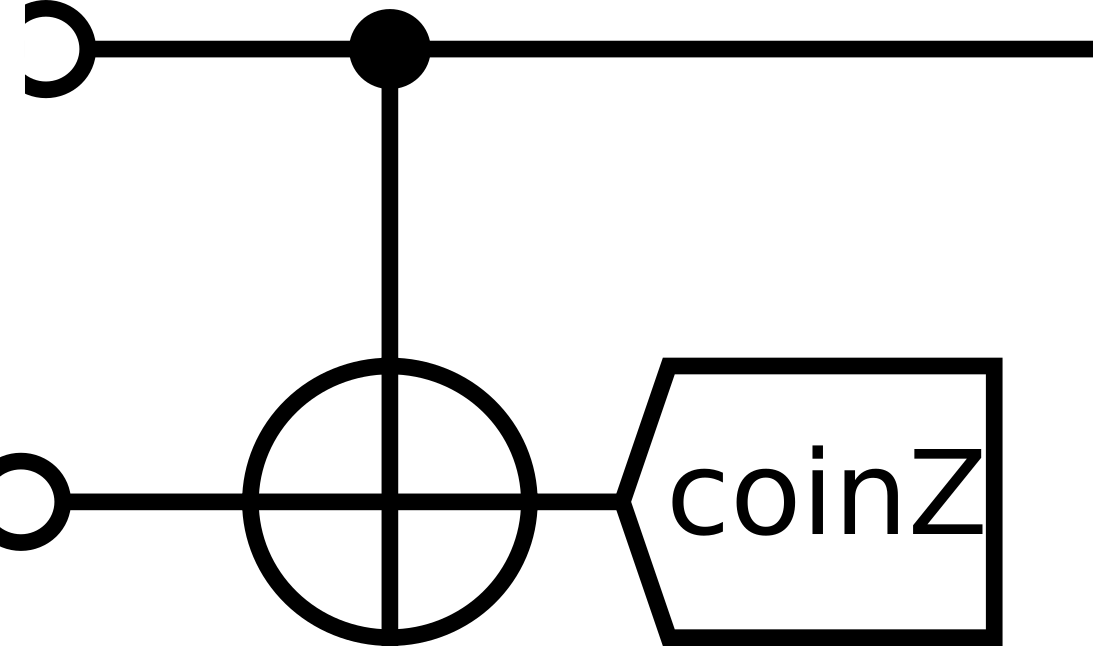}
\par\end{center}

We assume that Alice and Bob started with two perfect Bell pairs in
the state A. Each of the two registers (the one Alice uses to store
her two qubits and the one Bob uses for his) are subject to complete
depolarization with probability $\varepsilon=1-p_{2}$. This is equivalent
to saying that for each of the registers there is probability $\frac{\varepsilon}{16}$
for one of the 16 two qubit Pauli operators to be applied to the state.
Writing the possibilities down in a table (columns correspond to the
possible errors on the top/preserved qubit, rows correspond to the
possible errors on the bottom/sacrificial, and each cell gives the
corresponding tensor product):
\begin{center}
\begin{tabular}{|c|c|c|c|c|c|}
\cline{3-6} 
\multicolumn{1}{c}{} &  & \multicolumn{4}{c|}{{\footnotesize{}on preserved}}\tabularnewline
\cline{3-6} 
\multicolumn{1}{c}{} &  & I & X & Y & Z\tabularnewline
\hline 
\multirow{4}{*}{\begin{turn}{90}
{\footnotesize{}on sacrificial}
\end{turn}} & I & II & XI & YI & ZI\tabularnewline
\cline{2-6} 
 & X & IX & XX & YX & ZX\tabularnewline
\cline{2-6} 
 & Y & IY & XY & YY & ZY\tabularnewline
\cline{2-6} 
 & Z & IZ & XZ & YZ & ZZ\tabularnewline
\hline 
\end{tabular}
\par\end{center}

If we are to perform a coincidence measurement immediately, we will
be able to detect the errors that have occurred on the sacrificial
qubit, however they are not correlated with errors that have occurred
on the preserved qubit, therefore no errors on the preserved Bell
pair would be detected. However, if we perform a CNOT gate, errors
on either qubit will be propagated to the other one, and we will be
able to detect some of the errors that have occurred on the Bell pair
to be preserved by measuring the sacrificial Bell pair. Bellow we
describe how the errors propagate:
\begin{center}
\includegraphics[width=3cm]{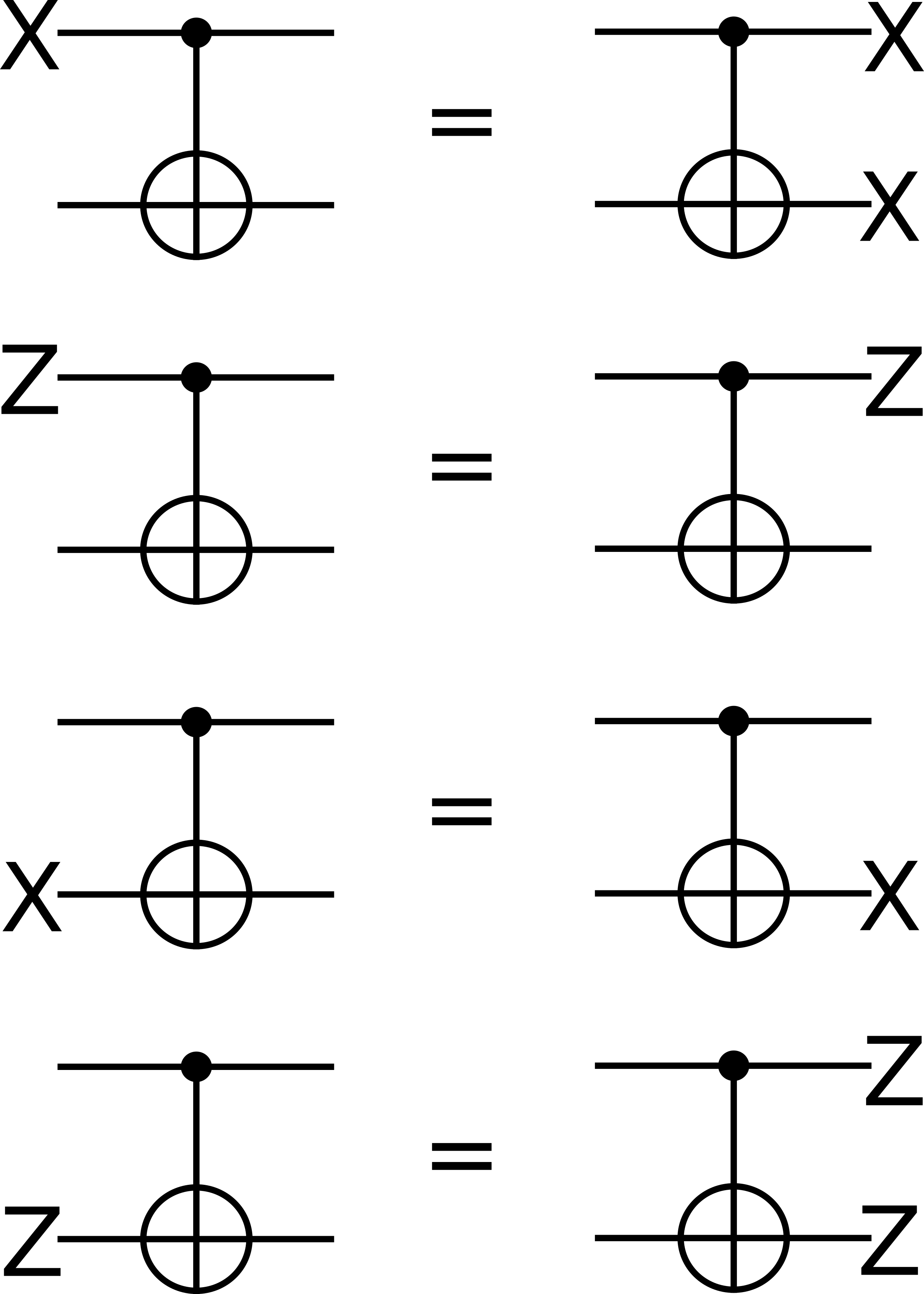}
\par\end{center}

After the CNOT gate we have the following redistribution of errors:
\begin{center}
\begin{tabular}{|c|c|c|c|c|c|}
\cline{3-6} 
\multicolumn{1}{c}{} &  & \multicolumn{4}{c|}{{\footnotesize{}on preserved}}\tabularnewline
\cline{3-6} 
\multicolumn{1}{c}{} &  & I & X & Y & Z\tabularnewline
\hline 
\multirow{4}{*}{\begin{turn}{90}
{\footnotesize{}on sacrificial}
\end{turn}} & I & II & XX & YX & ZI\tabularnewline
\cline{2-6} 
 & X & IX & XI & YI & ZX\tabularnewline
\cline{2-6} 
 & Y & ZY & YZ & XZ & IY\tabularnewline
\cline{2-6} 
 & Z & ZZ & YY & XY & IZ\tabularnewline
\hline 
\end{tabular}
\par\end{center}

Performing a coincidence Z measurement on the sacrificial Bell pair
will be able to detect X or Y errors, which leaves us with the following
table conditioned on successful measurement.
\begin{center}
\begin{tabular}{|c|c|c|c|c|c|}
\cline{3-6} 
\multicolumn{1}{c}{} &  & \multicolumn{4}{c|}{{\footnotesize{}on preserved}}\tabularnewline
\cline{3-6} 
\multicolumn{1}{c}{} &  & I & X & Y & Z\tabularnewline
\hline 
\multirow{4}{*}{\begin{turn}{90}
{\footnotesize{}on sacrificial}
\end{turn}} & I & II &  &  & ZI\tabularnewline
\cline{2-6} 
 & X &  & XI & YI & \tabularnewline
\cline{2-6} 
 & Y &  & YZ & XZ & \tabularnewline
\cline{2-6} 
 & Z & ZZ &  &  & IZ\tabularnewline
\hline 
\end{tabular}
\par\end{center}

Out of the 8 possibilities (16 initially), 2 (II \& IZ) are harmless
to the preserved Bell pair and the remaining 6 are damaging, which
leaves us with infidelity, to first order, $\frac{6}{16}\varepsilon\times2$
(the factor of 2 comes from the fact that both Alice and Bob are subject
to depolarization errors).

We can now augment the circuit with another level of detection that
will be able to detect the Z error on the sacrificial Bell pair:
\begin{center}
\includegraphics[height=3.5cm]{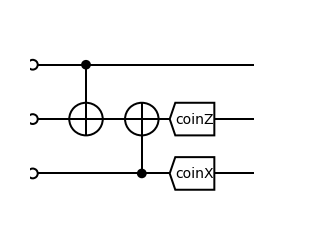}
\par\end{center}

To first order any errors contributed by this extension are negligible
and can not propagate back to the preserved Bell pair. The Z error
that might have occurred on the middle line (and was left undetected)
will now propagate to the bottom line and be detected by the coincidence
X measurement leaving us with the following table:
\begin{center}
\begin{tabular}{|c|c|c|c|c|c|}
\cline{3-6} 
\multicolumn{1}{c}{} &  & \multicolumn{4}{c|}{{\footnotesize{}on preserved}}\tabularnewline
\cline{3-6} 
\multicolumn{1}{c}{} &  & I & X & Y & Z\tabularnewline
\hline 
\multirow{4}{*}{\begin{turn}{90}
{\footnotesize{}on sacrificial}
\end{turn}} & I & II &  &  & ZI\tabularnewline
\cline{2-6} 
 & X &  & XI & YI & \tabularnewline
\cline{2-6} 
 & Y &  &  &  & \tabularnewline
\cline{2-6} 
 & Z &  &  &  & \tabularnewline
\hline 
\end{tabular}
\par\end{center}

Out of the 4 undetected errors, 3 still harm the preserved Bell pair,
so we are left with fidelity $\frac{3}{16}\varepsilon\times2$. Those
three are undetectable as they do not propagate to the sacrificial
qubits (they act as the identity on the sacrificial qubits). As such,
using bigger registers (wider circuits) would provide only small higher-order
corrections.

\begin{figure}
\includegraphics[width=8cm]{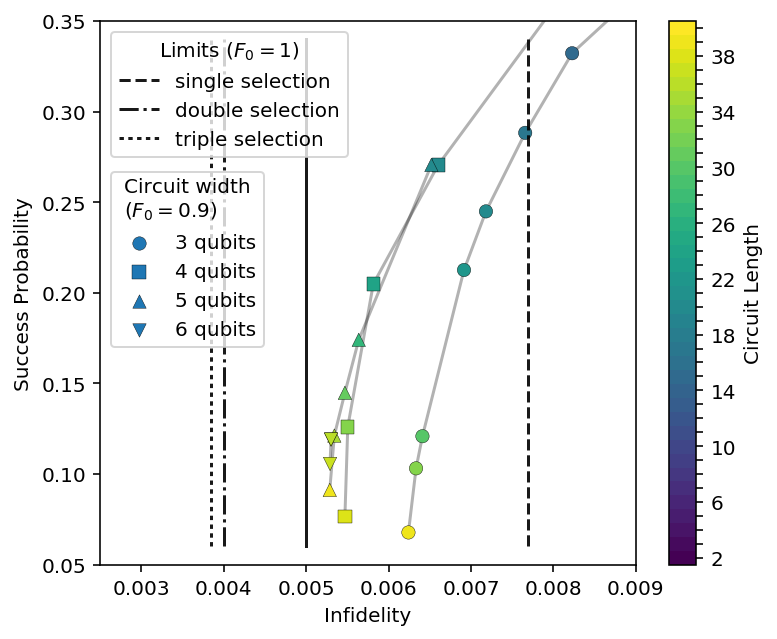}

\caption{\label{fig:N3vsN4-b}Same as \ref{fig:N3vsN4} but we add some additional
information. Dashed vertical lines, corresponding to perfectly initialized
($F_{0}=1$, $p_{2}=\eta=0.99$) short purification circuits, are
shown as a guide to how well the circuits perform in terms of initialization
versus operational errors as described in the main text. In ``single
selection'' circuits each party uses registers of size two, size
three for ``double selection'', and size four for ``triple selection''.}
\end{figure}

Finally, the asymptote reached by our circuits contains one additional
source of infidelity. The black vertical lines in Fig.~\ref{fig:N3vsN4-b}
correspond to short circuits with zero initialization error. However,
a real purification protocol would need multiple rounds of purification
until it lowers the non-zero initialization error to the steady-state
floor governed by the operational error. In this steady state an additional
round of purification would be able to detect only 2 of the possible
3 Pauli error that were already present, therefore raising the bound
of the achievable infidelity from $\frac{3}{16}\varepsilon\times2$
to $\frac{3+1}{16}\varepsilon\times2$ (to first order in $\varepsilon$).
For the parameters of Fig.~\ref{fig:N3vsN4-b} this would correspond
to an asymptote at infidelity of $0.005$ which is indeed what we
observe.

The vertical lines of Fig.~\ref{fig:N3vsN4-b} are slightly offset
from the values quoted above because we used exact numerics for the
plots, as opposed to the first-order calculations of this section.

\section{Shortest multi-pair purification circuits}

In the main text we introduced circuits to be used as benchmarks of
initialization-vs-operation errors. The idea was to show what performance
is provided by a circuit applied to perfectly initialized raw Bell
pairs, or in other words, how much damage is caused by operation errors
if we start with perfect initialization (as done in Fig.~\ref{fig:N3vsN4}
and Fig.~\ref{fig:N3vsN4-running-p}). To do that we found with brute-force
enumeration the best ``short'' circuits, i.e. circuits that do not
reinitialize any of the consumed Bell pairs. They are named in the
manner introduced in \citep{fujii2009entanglement}. The triple select
circuit is actually a generalization of the circuit from \citep{fujii2009entanglement}.
As described in \citep{fujii2009entanglement} and in the main text
of our manuscript, double selection significantly outperforms single
selection, and extending the double selection circuit to a triple
selection circuit provides only modest higher-order improvements.

\begin{figure}
\begin{centering}
\includegraphics[height=2.5cm]{double_sel}\includegraphics[height=3cm]{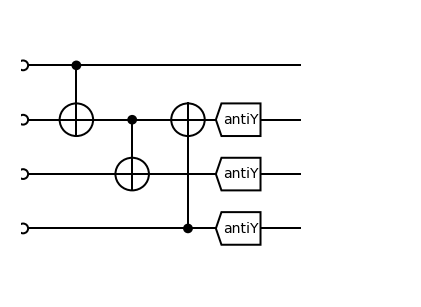}
\par\end{centering}
\caption{\label{fig:double-triple-selection}A double selection circuit on
the left and a triple selection circuit on the right. We show only
Bob's side of the circuit. The circuit from Fig.~\ref{fig:simple-purification}
can be referred to as a single selection circuit. As explained in
the main text, there are many circuits with equivalent performance,
related to the given circuits by permutation of the Bell basis.}
\end{figure}

\section{More about initialization errors, operational errors, and the length
of the circuit}

In Fig.~\ref{fig:N3vsN4} the vertical lines showed the ``operational
error'' limit which one would reach if there were no initialization
errors. 

\begin{figure}
\includegraphics[width=7cm]{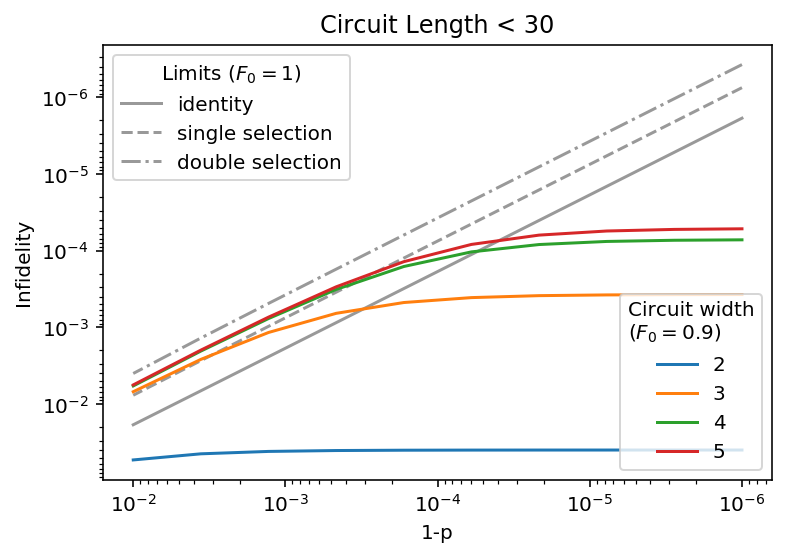}

\includegraphics[width=7cm]{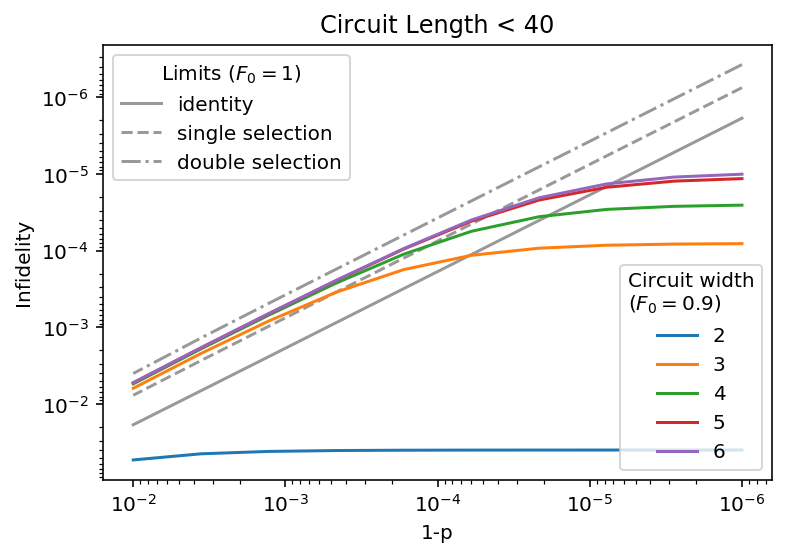}

\caption{\label{fig:N3vsN4-running-p}For each family of generated circuits
of various width (color) and for a given operational infidelity (x
axis) we show the best achievable final Bell pair fidelity by a circuit
in that family. The top plot limits the permitted circuits to length
less than 30 operations, and there is a limit of less than 40 operations
for the bottom plot. Three important observations can be made: (1)
as long as we can lower the operational error, we can design a long
enough circuit that is not affected by the initialization error; (2)
a wider register outperforms smaller registers and reaches a point
of diminishing returns at smaller operational errors; (3) as already
mentioned, circuits of width 2 are insufficient for arbitrary suppression
of the initialization error as they detect only 2 of the possible
3 single-qubit errors (X, Y, and Z). The grey lines follow the same
conventions as in Fig.~\ref{fig:N3vsN4} - short perfectly initialized
circuits used as a benchmark. The triple selection circuit from Fig.~\ref{fig:N3vsN4}
is omitted as it can not be distinguished from the double select circuit
on this scale. The ``identity'' line corresponds to what would happen
if we simply depolarize a single perfect Bell pair with probability
$1-p$.}
\end{figure}

To make the comparison between initialization and operational errors
clearer we provide Fig.~\ref{fig:N3vsN4-running-p} which drops the
``success probability'' axis of Fig.~\ref{fig:N3vsN4} and instead
shows how the performance varies with $p_{2}$. In it one can see
that the initialization infidelity is not a limiting factor - as long
as the operational infidelity can be lowered, we can find longer circuits
that iteratively get rid of the initialization error. For a sufficiently
long circuit we reach a point of saturation, where, as described above,
the operational error in the last operation dominates the infidelity
of the final Bell pair. Similarly, if the circuit is wider (i.e. the
register is bigger) we can obtain higher final fidelities for a fixed
operational error, and the point saturation occurs at even lower operational
error levels.

\section{Circuits for registers with dedicated communication qubit}

\begin{figure}
\includegraphics[width=6.69cm]{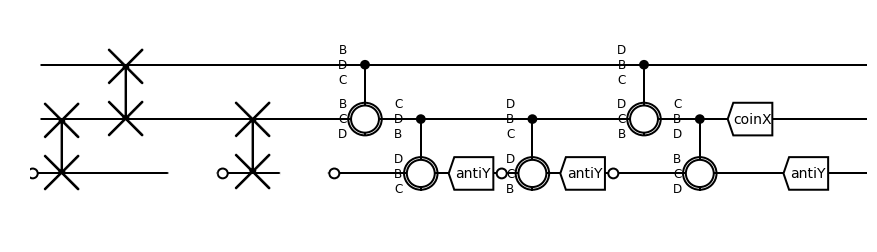}

\includegraphics[width=8cm]{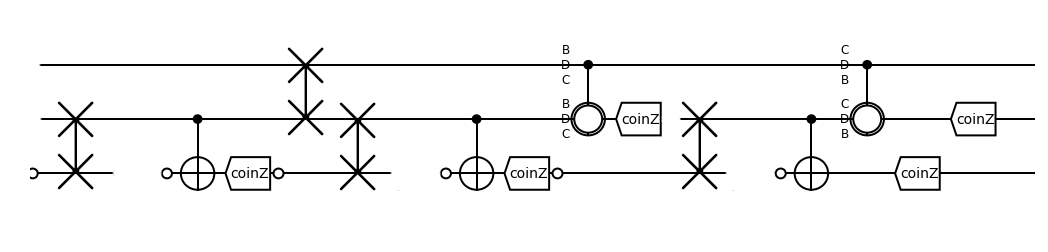}

\caption{\label{fig:hotcold}The top circuit is the one produced by our modified algorithm, while the bottom one is
from Nigmatullin et al.~\citep{nigmatullin2016minimally}. We use the same notation as in the rest of the manuscript, in
particular the small circles represent the reestablishment of a raw Bell pair (which here is only possible on the bottom-most qubit,
i.e. the communication qubit). Our circuit requires only 5 raw Bell pairs (as opposed to 6 for the circuit
from~\citep{nigmatullin2016minimally}), produces a final Bell pair of infidelity of 1.77\% (as opposed to 2.46\%), at a
success rate of 48\% (as opposed to 43\%). The optimization and evaluation was done for $F_{0}=0.9$, $p_{2}=\eta=0.99$, without
memory errors or errors from single-qubit unitaries. As discussed in the rest of the text, another circuit could be
optimal for a different error model.}
\end{figure}

Some hardware implementations of quantum registers have constraints on the two-qubit operations they can perform.
For instance, only a single "communication" qubit might be able to establish
a raw Bell pair with the remote register. Works like \citep{nigmatullin2016minimally} suggest purification circuits
specifically designed to fulfill such constraints. To see how our optimization protocol compares to such hand-crafted
circuits we made the following modifications:
SWAP gates were added to the set of permitted operations;
two-qubit gates were made available only on nearest neighbors;
measurements on anything but the communication qubit became destructive;
a reset operation (production of a new raw Bell pair) was added as a possibility, but only for the communication qubit.
For the error regime we considered we were able to find a shorter circuit with higher fidelity and similar success
probability as seen in Fig.~\ref{fig:hotcold}.

\section{Different results when optimizing in different regimes}

\begin{figure}
\includegraphics[width=7cm]{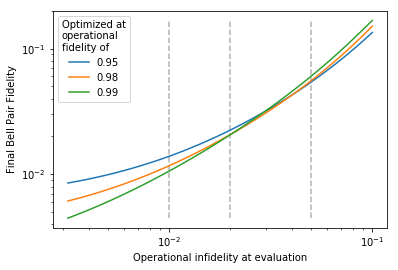}

\caption{\label{fig:compare-regime}Each colored line corresponds to the result
of a circuit optimization ran the given $p=p_{2}=\eta$ (depolarization
noise). Then each of these circuits are evaluated at various values
of $p$, different from the one they were optimized for. The x axis
is the value of $1-p$ at evaluation. The y axis is the obtained value
of $1-F$. The dashed lines mark the values at which the optimizations
were ran. The length of the circuits was constrained to $\le12$,
the width was 3 and $F_{0}=0.9$.}
\end{figure}

\begin{figure}
\includegraphics[width=7cm]{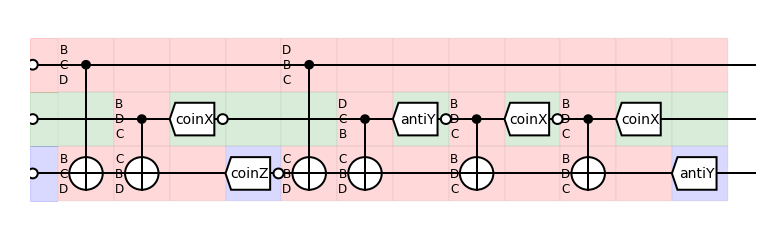}

\includegraphics[width=6.46cm]{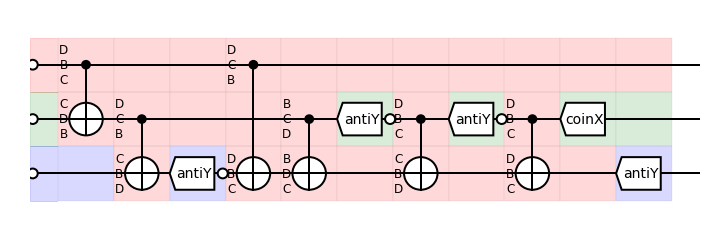}

\includegraphics[width=7cm]{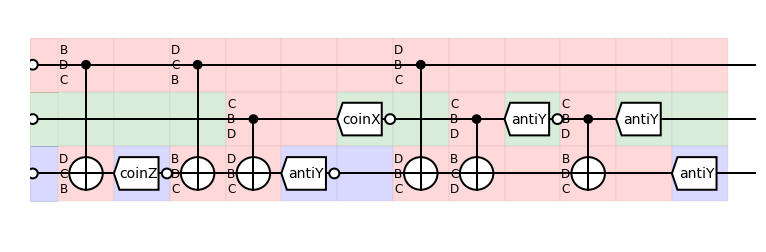}\caption{Top to bottom, the 0.95, 0.98, and 0.99 circuits from Fig.~\ref{fig:compare-regime}}
\end{figure}

In Fig.~\ref{fig:compare-regime} we demonstrate the importance of
optimizing your purification circuit for the exact hardware at which
it would be ran. One can see that each of the three circuits outperforms
the other two, only within a small interval around the parameter regime
for which it was optimized.

\section{Infidelity axes}

\begin{figure}
\begin{center}
\includegraphics[width=7cm]{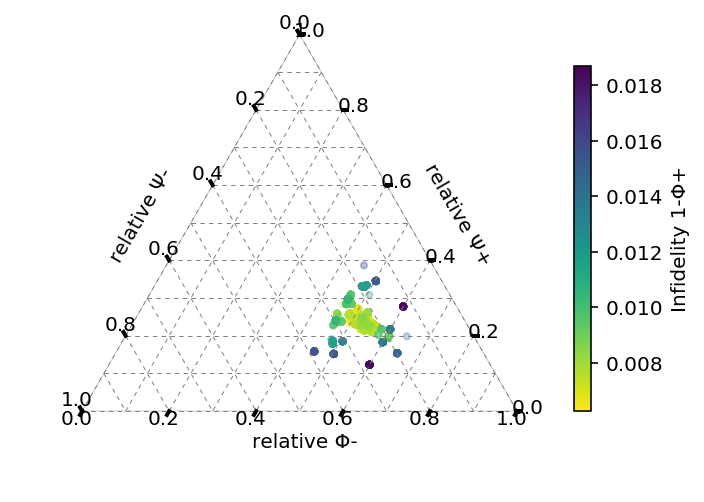}

\includegraphics[width=7cm]{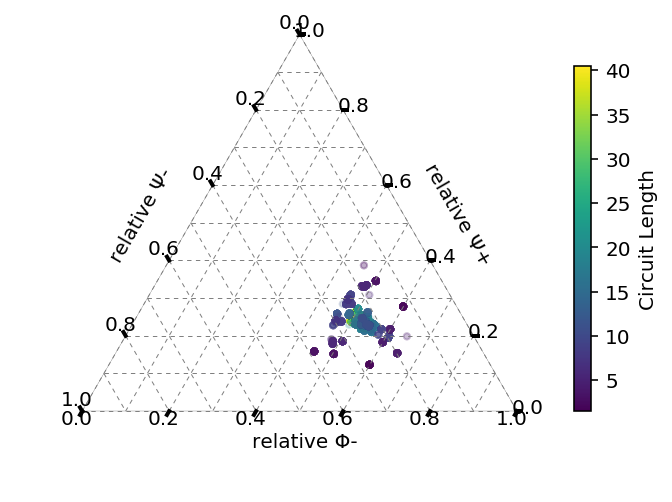}
\end{center}
\caption{\label{fig:ternary}Each point corresponds to one of the circuits
shown in Fig.~\ref{fig:comparing-circuits}. In the top plot they
are colored by the final infidelity of the Bell pair produced by the
circuit. In the bottom is the same plot, but the color corresponds
to the length of the circuit. The 3 axes of the ternary plot correspond
to the 3 components of the infidelity. As ternary plots require the
3 coordinates to fulfill a constraint, we plot the relative infidelity.
For instance, the left axis (corresponding to the height of a point
in the plot) shows the probability to be in $\psi_{+}$ divided by
the total probability to be in a state different from $\phi_{+}$
(the state being purified). Being in the center of the triangle means
the infidelity in the final result is pure depolarization. Being in
one of the corners means that one of the infidelity components dominates
the other two. Being near the midpoint of a side means one of the
infidelity components is much smaller than the other two. The 6 symmetries
of this triangle correspond to the six permutations of $\{B,C,D\}$.}
\end{figure}

Even if the error model for the circuits and initialization is the
depolarization model, the error in the final result of the purification
needs not be depolarization. The infidelity in the final result has
three components - the probabilities to be in states $\psi_{-}$,
$\psi_{+}$, and $\phi_{-}$, respectively. Different purification
circuits affect the three infidelity components differently, and giving
different weights in the cost function of the optimization algorithm
might be important, depending on the goal (for instance, if the purified
Bell pairs are used for the creation of a GHZ state, the particular
implementation might be more susceptible to phase errors, in which
case that component would be assigned a higher weight). In Fig.~\ref{fig:ternary}
we show the distribution of the infidelity components of the purified
Bell pair for each of the circuits we have generated. Of note is that
longer circuits reach nearer the pure depolarization error, by virtue
of lowering the infidelity to the level of diminishing returns where
the depolarization from to the final operation dominates. Moreover,
the results are biased to one particular type of error, due to the
particular choice of ``genes'' described in the main text, namely,
the $\{B,C,D\}$ permutations are performed before the CNOT gate and
measurements. This bias can be removed if necessary by applying one
final $\{B,C,D\}$ permutation (the six $\{B,C,D\}$ permutations
correspond to the six symmetries of the triangle in Fig.~\ref{fig:ternary}).

\section{Implementation of the various permutation operations}

Here we give explicitly what single-qubit Clifford operations are
necessary in order to perform a permutation of the Bell basis. $H$
stands for the Hadamard gate and $P$ states for the phase gate (in
parenthesizes we mark whether the permutation is a rotation or a reflection
of the triangle).
\begin{center}
\begin{tabular}{|c|c|c|}
\hline 
permutation & Alice does & Bob does\tabularnewline
\hline 
\hline 
BCD & nothing & nothing\tabularnewline
\hline 
BDC(refl) & $H$ & $H$\tabularnewline
\hline 
DCB(refl) & $HPH$ & $PHP$\tabularnewline
\hline 
CDB(rot) & $PH$ & $(HP)^{2}$\tabularnewline
\hline 
DBC(rot) & $(PH)^{2}$ & $(HP)^{4}$\tabularnewline
\hline 
CBD(refl) & $H(PH)^{2}$ & $H(HP)^{4}$\tabularnewline
\hline 
\end{tabular}
\par\end{center}

Even though the decomposition of these operations in terms of H and
P has different lengths, in practice these operations are equally
easy to implement on real hardware.

\section{Canonicalization of generated circuits}

Many redundancies can appear in the population of circuits subjected
to simulated evolution. To simplify the analysis of the results we
filter the generated circuits by first ensuring that for each circuit:
\begin{itemize}
\item the first operation is not a measurement;
\item does not contain two immediately consecutive measurements on the same
qubit;
\item does not contain unused qubits;
\item does not contain measurement and reset of the top-most qubit pair
(the one containing the Bell pair to be purified);
\item does not contain non-measurement as a last step;
\end{itemize}
and then for the non-discarded circuits we perform the following canonicalization:
\begin{itemize}
\item reorder the qubits of the register so that the qubit closest to the
top-most qubit is the one to be measured last, the second closest
is measured second to last, etc;
\item if a two-qubit gate and a measurement commute (i.e. they affect different
qubits of the register), reorder them so that the gate is always before
the measurement (in an implementation of that circuit, those two operations
will be executed in parallel);
\item if two two-qubit gates affect different qubits, put the one that affects
top-most qubits before the one that affects lower qubits (in an implementation
of that circuit, those two operations will be executed in parallel).
\end{itemize}
The canonicalization rules are arbitrary and any other consistent
set can be used, at the discretion of the software writer. However,
they ensure that two circuits that are physically equivalent are not
presented multiple times in the final result, substantially lowering
the circuits that need to be evaluated. The set described above is
not exhaustive, as there are other, more complex, equivalences that
we have not considered.

\begin{figure*}[t]
\begin{center}
\includegraphics[width=13cm]{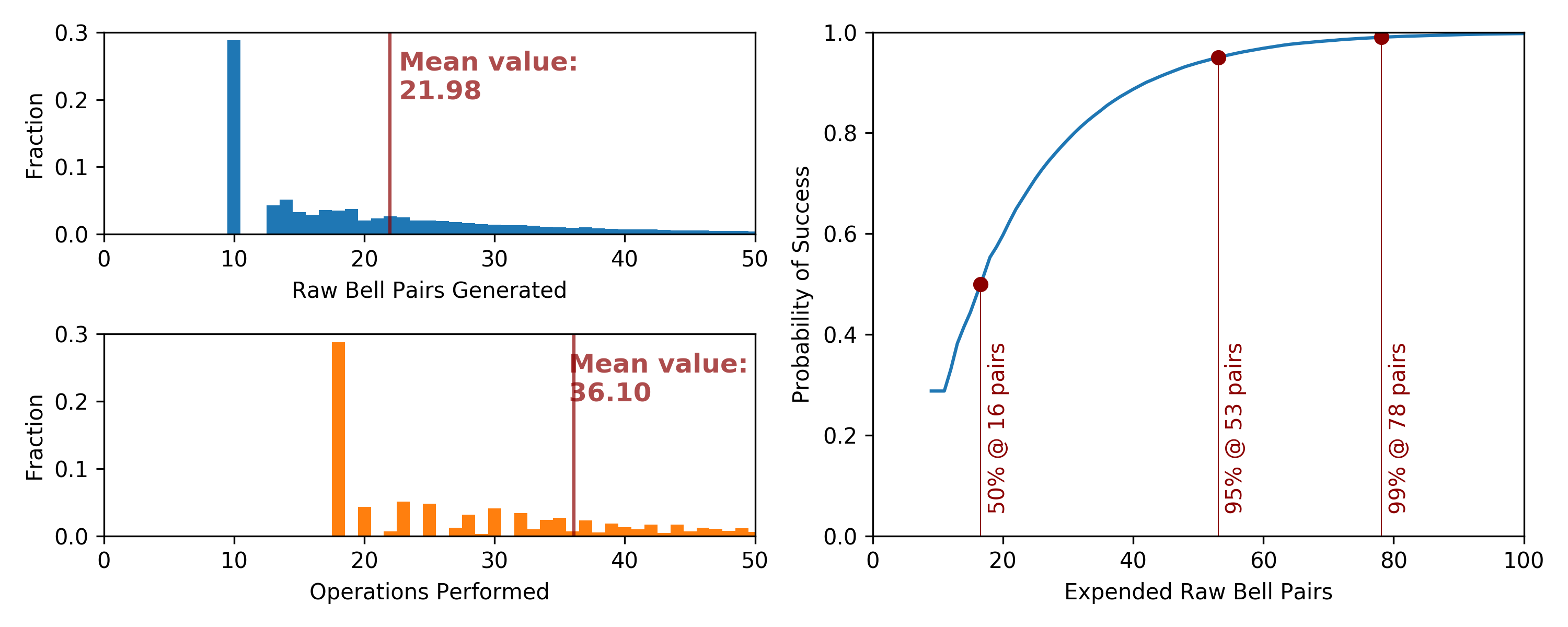}
\end{center}
\caption{\label{fig:monte-carlo}Monte Carlo evaluation of overhead due to
restarts of failed measurements. The evaluation is from the circuit
from Fig.~\ref{fig:L17-STRINGENT} at $F_{0}=0.9$ and $p_{2}=\eta=0.99$.
On the left we have the histogram of completed runs in terms of how
many operations a run takes to successfully complete. In each histogram,
the mean value of the distribution is showed as well. On the right
we have the probability for the protocol (in which reinitializations
are permitted) to successfully complete in terms of how many Bell
pairs were used (i.e. it is the cumulative version of the top-left
plot).}
\end{figure*}

\begin{figure}
\includegraphics[width=7cm]{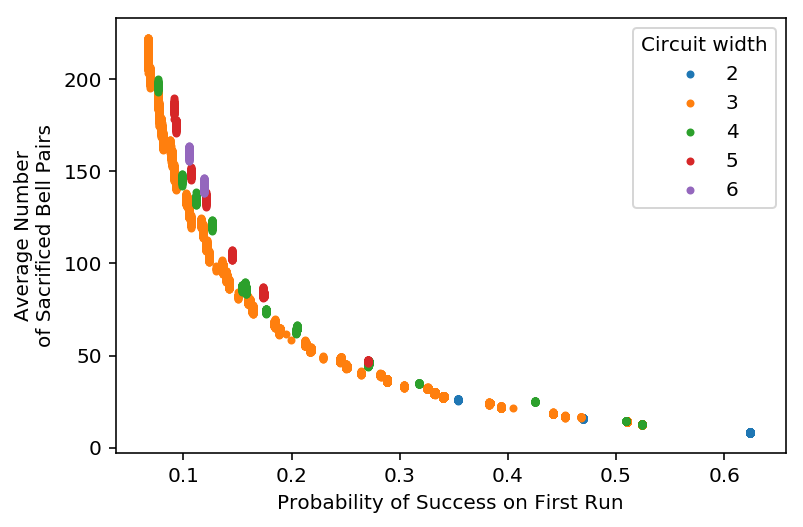}\caption{\label{fig:yield}The relationship between overhead and success probability
for the designs generated by our algorithm. Longer circuits have both
lower success probability and higher overhead (expended Bell pairs).
However, as shown in the rest of manuscript, longer circuits approach
asymptotically the upper bound of performance.}
\end{figure}

\begin{figure}
\begin{center}
\includegraphics[width=7cm]{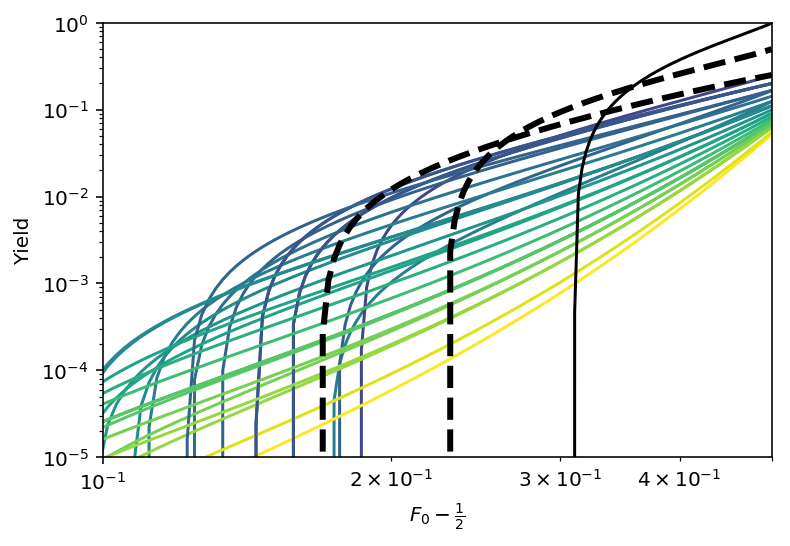}

\includegraphics[width=7cm]{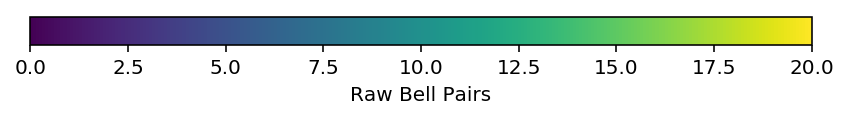}
\end{center}
\caption{\label{fig:yield_hashing}Plotted is the "hashing yield" of a given purification protocol versus the initial fidelity of the raw Bell pairs.
The solid black line is the yield of the hashing method on its own. The dashed lines are the yields of the Oxford method (either on its own or used recursively)
used as initial purification step and then continued by the hashing method. The colored lines are lower bounds for various protocols we have generated,
used as initial purification step and then continued by the hashing method. The color represents the upper bound of how many 
raw Bell pairs the protocol requires. Of note is that although the hashing method on its own is the highest yield protocol when $F_{0}\approx 1$,
around $F_{0}\approx 0.81$ we get $1-H(F_0)=0$ which requires the preprocessing (concatenation) steps, to bring the intermediate stage fidelity to a workable level. We do not optimize for the hashing yield, given that
it is defined for perfect local gates and measurements.}
\end{figure}

\section{Yield in the absence of measurement and operational errors}

As mentioned in the main text, if we concatenate a finite circuit with the hashing method, we can define a "hashing" yield
even for our finite circuits. It is defined as
$\frac{P}{N}(1-H(F))$, where $P$ is the success probability for the circuit under consideration, $N$ is the number of raw Bell pairs being
sacrificed, $F$ is the fidelity of the Bell state produced by the circuit, and $H$ is the entropy in that Bell state (the information
in that Bell state is the resource exploited by the hashing method to asymptotically reach unit fidelity, and consequently, $1-H(F_0)$, where
$F_0$ is the initial raw Bell pair fidelity, would be the yield one can achieve when using only the hashing method).
This "hashing yield" definition faces a couple of problems in the parameter regime we work with. First, this quantity only refers
to yield for circuits devoid of measurement and operational errors, but as we have seen, we need to consider such errors
because the best finite purification circuit depends on the error model. Moreover, as our circuits might fail early, before
all of the raw Bell pairs have been engaged, the factor $\frac{1}{N}$ is only a worst case bound. As such, the yield given by this
expression is only a lower bound. We compare the yields given by the circuits we have generated to a number of known circuits in
Fig.~\ref{fig:yield_hashing} (in the absence of measurement and operational errors). However, we remind the reader that the
yield is not the cost function in our optimization algorithms, as our circuits are specifically optimized to deal with the
aforementioned local errors.

\section{Analytical expressions for the final fidelity}

Our software also produces a symbolic analytical expression for the
fidelities obtained by each circuit. The quality of the raw Bell pairs
is expressed as the quadruplet of probabilities to be in each of the
Bell basis states $(F_{0},q,q,q)$ where $q=\frac{1-F_{0}}{3}$ (more
general non-depolarization error models are available as well). The
purification circuit acts as a map that takes $(F,q,q,q)$ to the
quadruplet $(F_{A},b,c,d)$ representing the probabilities that the
final purified Bell pair is in each of the Bell basis states.

The permutations of the Bell basis (i.e. all the local Clifford operations
we are permitting) are polynomial maps, i.e. the output quadruplet
contains polynomials of the variables in the input quadruplet. Depolarization
is a polynomial map as well. Measurement without normalization is
a polynomial map, but it becomes a rational function if normalization
is required.

By postponing normalization until the very last step, we can use efficient
symbolic polynomial libraries like Sympy (using generic symbolic expressions
is much slower than using polynomials). The final result is given
as a series expansion of the normalized expression (in terms of the
small parameters $1-F_{0}$, $1-p_{2}$, and $1-\eta$).

\section{Monte Carlo simulations of restart overhead}

As mentioned in the main text, one needs to consider how a protocol
proceeds when a measurement fails. If there is a subcircuit that can
be restarted, one needs not redo the entire protocol, rather only
reinitialize at the point where the subcircuit starts. However, if
the top-most qubit pair, the one holding the Bell pair, is entangled
with the qubit pair that undergoes a failed coincidence measurement,
then the entire protocol has to be restarted. For most of our circuits,
such subcircuits do not exist, but they are common among manually
designed circuits. Our software automatically finds subcircuits and
runs a Monte Carlo simulation of the sequence of measurements and
reinitializations in order to evaluate the average resource usage
as shown in Fig.~\ref{fig:monte-carlo}.

The overhead estimated this way also proves to be closely related
to the success probability of the given protocol, to be expected given
that greater overhead implies more imperfect operations which implies
higher chance of a fault (Fig.~\ref{fig:yield}).

\end{document}